\documentclass[a4paper,11pt]{article}

\usepackage{amsmath}
\usepackage{amssymb}
\usepackage{multirow}
\usepackage[textsize=tiny]{todonotes}
\usepackage[english]{babel}
\usepackage{xcolor}
\usepackage{graphicx}
\usepackage{url}
\usepackage{natbib}
\usepackage{color}
\usepackage[normalem]{ulem}

\newcommand{\eg}[0]{\textit{e.g.}}
\newcommand{\ie}[0]{\textit{i.e.}}
\newcommand{\ignore}[1]{}

\newcommand{\ELneighX}[0]{M}
\newcommand{\ELneigh}[1]{\ELneighX(#1)}

\newcommand{\barriers}{\texttt{barriers}}
\newcommand{\lid}{\texttt{lid}}

\newcommand{\FPT}{\tau{}}

\newcommand{\change}[1]{#1}

\title{Memory efficient RNA energy landscape exploration}

\author{Martin Mann\,$^1$\footnote{to whom correspondence
should be addressed: \url{http://www.bioinf.uni-freiburg.de}}, Marcel
Kuchar\'ik\,$^2$, Christoph Flamm\,$^2$\\ and Michael T. Wolfinger\,$^{2,3,4}$}

\date{}

\begin{document}
\maketitle{}

\begin{center}
{\footnotesize $^1$~Bioinformatics Group, University of Freiburg,
  Georges-K\"ohler-Allee 106, D-79110 Freiburg, Germany \\$^2$~Institute for
  Theoretical Chemistry, University of Vienna, W\"{a}hringerstra{\ss}e 17, 1090
  Vienna, Austria\\ $^3$~Center for Integrative Bioinformatics Vienna (CIBIV),
  Max F. Perutz Laboratories, University of Vienna \& Faculty of Computer
  Science, University of Vienna, Dr.~Bohr-Gasse 9, 1030 Vienna, Austria.\\
  $^4$~Department of Biochemistry and Molecular Cell Biology, Max F. Perutz
  Laboratories, University of Vienna, Dr.~Bohr-Gasse 9, 1030 Vienna, Austria\\}
\end{center}

\begin{abstract}
 Energy landscapes provide a valuable means for
studying the folding dynamics of short RNA molecules in detail by modeling
all possible structures and their transitions. Higher abstraction levels
based on a macro-state decomposition of the landscape enable the study of
larger systems, however they are still restricted by huge memory
requirements of exact approaches.

We present a highly parallelizable local enumeration
scheme that enables the computation of exact macro-state transition models
with highly reduced memory requirements. The approach is evaluated on RNA
secondary structure landscapes using a gradient basin definition for
macro-states. Furthermore, we demonstrate the need for exact transition
models by comparing two barrier-based appoaches and perform a detailed
investigation of gradient basins in RNA energy landscapes.

Source code is part of the C++
Energy Landscape Library available at
\url{http://www.bioinf.uni-freiburg.de/Software/}.

\end{abstract}

\section{Introduction}

The driving force of disordered systems in physics, chemistry and biology
is characterized by coupling and competing interaction of microscopic
components. At a qualitative level, this is reflected by the potential
energy function and often results in complex topological properties induced
by individual conformational degrees of freedom. It seems fair to say that
it is practically impossible to compute dynamic and thermodynamic
properties directly from the Hamiltonian of such a complex system. However,
analyzing the underlying energy landscape and its features directly
provides a valuable alternative.

\change{Here, we focus on RNA molecules and their folding kinetics. RNAs are key
players in cells acting as regulators, messengers, enzymes, and many more roles.
In many cases, a specific structure is crucial for biological specificity and
functionality. The formation of these functional structures, \ie{} the folding
process, can be studied at the level of RNA energy
landscapes~\citep{Flamm:08a,Geis:08a}.

RNA is composed of the biophysical alphabet $\{$\texttt{A,C,G,U}$\}$ and
has the ability to fold back onto itself by formation of discrete base
pairs, thus forming secondary structures. The latter provide a natural
coarse-graining for the description of the thermodynamic and kinetic
properties of RNA, because, in contrast to proteins, the secondary
structure of RNA captures most of the folding free energy. This is
accomodated by novel approaches for predicting three-dimensional RNA
structures from secondary structures \citep{Popenda:2012}.

Formally, an RNA secondary structure is defined as a set of base pairs
between the nuclear bases complying with the rules: (a)~only \texttt{A-U},
\texttt{G-C}, and \texttt{G-U} pairings are allowed, (b)~any base is
involved in maximal one base pair, and (c)~the structure is nested, \ie{}
there are no two base pairs with indices $(i,j), (k,l)$ with
$i<k<j<l$. Summation over the individual base pair binding energies and entropic
contributions for unpaired bases defines the energy function $E$
\citep{Hofacker:94a,Tinoco:71,Freier:86}. The degeneracy of this energy
definition is countered via a structure ordering based on their string
encoding~\citep{Flamm:02}.
We refer to the literature~\citep{Flamm:08a,Chen:08} for more details.}

\change{In this work, we study the folding kinetics of RNA molecules by
means of a discrete energy landscape approach.} While stochastic folding
simulations based on solving the Master equation are limited to relatively
short sequence lengths~\citep{Flamm:00a,Aviram:12}, a common approach to
studying biopolymer folding dynamics is using a coarse grained model that
partitions the energy landscape into distinct basins of attraction, thus
assigning macro-states to each basin\change{~\citep{Wolfinger:04a}. The
basin decomposition and computation has been described in different
contexts, including Potential Energy Landscapes~\citep{Heuer:08}, RNA
kinetics~\citep{Flamm:08a} and lattice protein
folding~\citep{Wolfinger:06a,Tang:2012}.} Given appropriate transition
rates between macro-states (optionally comprised of rates among
micro-states that form a macro-state), the dynamics can be modeled as
continuous-time Markov process and solved directly by numerical
integration~\citep{Wolfinger:04a}. While suitable for system sizes up to
approx. 10,000 states, improvements to this approach are currently subject
to our research, allowing investigation of up to a few hundred throusand
states by incorporating sparsity information and additional approximations.

The crucial step in the procedure sketched above is to obtain the transition
rates between macro-states. Global methods for complete~\citep{Flamm:02} or
partial~\citep{Sibani:99,Kubota05,Wolfinger:06a} enumeration of the energy
landscape are not applicable to large systems due to memory restrictions. On the
other side, sampling with high precision requires long sampling times
\citep{Mann:Klemm:11}. Therefore approximating the energy landscape by a subset
of important local minima, gained via sampling approaches or spectroscopic
methods \citep{Fuertig:2007,Aleman:2008,Rinnenthal:2011}, and transition
paths between them \citep{Noe:2008} has been investigated in the past
\change{\citep{Tang:05,Tang:08,Kucharik:2014}}.

We propose a novel, highly parallelizable and memory efficient local
enumeration approach for computing exact transition
probabilities. \change{While the method is intrinsically generic and can be
readily applied to other discrete systems, we exemplify the concept in
the context of energy landscapes of RNA secondary structures, based on
the Turner energy model \citep{xia:98}, as implemented in the
\texttt{Vienna RNA Package} \citep{Hofacker:94a,Lorenz:11a} and the Energy
Landscape Library \citep{Mann_ELL_BIRD07}. We evaluate the memory
efficiency and dynamics quality for different RNA molecules and report
features of gradient basin macro-states in RNA energy landscapes.}

\section{Discrete Energy Landscapes}

In the following, we will define energy landscapes for two levels of
abstraction: the \textit{microscopic level} covers all possible (micro-) states
of a system and its dynamics, while the \textit{macroscopic level} enables a
more coarse grained model of the system's dynamics, based on a partitioning of
all micro-states into macro-states. The macroscopic view is required when
studying the dynamics of larger systems.

\subsection{Microscopic Level}

Discrete energy landscapes are defined by a triple $(X,E,\ELneighX)$ given a
finite set of (micro-)states $X$, an appropriate energy function $E:X
\rightarrow {\mathbb R}$, and a symmetric neighborhood relation $\ELneighX : X
\rightarrow \mathcal{P}(X)$ (also known as move set), where $\mathcal{P}(X)$ is
the power set of $X$. The neighborhood $\ELneighX(x)$ is the set of all
neighboring states that can be directly reached from state $x$ by a simple move
set operation.

\change{Consequently, RNA energy (folding) landscapes can be defined at the
level of secondary structures, which represent the
micro-states $x \in X$. An RNA structure $y$ is neighbored to a structure $x$
($y\in\ELneigh{x}$) if they differ in one base pair only. While alternative
move set definitions are possible~\citep{Flamm:00a}, they are not considered in
this work for simplicity.}

Within this work, we consider time-discrete stochastic dynamics based on
Metropolis transition probabilities $p$ at inverse temperature~$\beta$:
\begin{eqnarray} 
p_{x \rightarrow y} &=& \Delta^{-1} \min\{ \exp( -\beta[E(y)-E(x)]),1\} 
\nonumber\\
&=& \Delta^{-1} \min\{ w(y)/w(x) , 1 \} \label{eq:pxy} \\
\text{with }w(x) &=& \exp({-\beta E(x)}) \\
\text{and }\Delta &=& \max_{x \in X}|\ELneighX(x)|.
\end{eqnarray} 
$w(x)$ is the Boltzmann weight of $x$. \change{Normalization is performed
  via the constant} $\Delta$\change{, which} is the maximal\change{ly
  possible} number of neighbors\change{/transitions} of any state. The
transition probability $p_{x \rightarrow y}$ is only defined for
neighboring states, \ie{} $y\in\ELneighX(x)$.

\subsection{Macroscopic Level}

Although desirable, studying dynamic properties at the microscopic level is
often not feasible due to the vastness of the state space $X$, even for
relatively small systems. An alternative approach is coarse graining \ie{}
lumping many micro-states into fewer macro-states, such that the
microscopic dynamics is resembled as closely as possible
\change{\citep{Wolfinger:04a}}.

This can be achieved by partitioning of the state space $X$ with a mapping
function $F : X \rightarrow B$ that uniquely assigns any micro-state in $X$
to a macro-state in $B$. With \change{$F^{-1}(b)$} we denote the inverse
function that gives the set of all $F$-assigned states for a macro-state
\change{$b\in B$}. Following
\change{\citep{Kramers:40,Wolfinger:04a,Flamm:08a,Mann:Klemm:11}}, we will
use the simplifying assumption that the probability of the system to be in
micro-state $x$ while it is in macro-state $b\in B$ is given by
\begin{eqnarray}
P_b(x) &=& \left\{ \begin{array}{ll}
w(x) Z_b^{-1} & \textrm{if } x \in F^{-1}(b) \\
0        	  & \textrm{otherwise}
\end{array} \right. \label{eq:pb} \\
\text{with }Z_b &=& \sum_{y \in F^{-1}(b)} w(y). \label{eq:zb}
\end{eqnarray}

Based on this, we can define the macroscopic transition probabilities $q_{b
\rightarrow c}$ between macro-states $b,c\in B$ by means of the
microscopic probabilities $p$ from Eq.~\ref{eq:pxy} as follows:
\begin{eqnarray} \label{eq:qbc}
q_{b \rightarrow c} 
&=& \sum_{x \in F^{-1}(b)} \left( P_b(x)
\sum_{y \in \ELneigh{x} \cap F^{-1}(c)} p_{x \rightarrow y}\right)
\label{eq:q_macro_1}\nonumber\\ 
&=& \sum_{(x,y)} P_b(x)\; p_{x \rightarrow y} \nonumber\\
&=& \sum_{(x,y)} \frac{w(x)}{Z_b}
\Delta^{-1} \min \{w(y)/w(x),1\}) \nonumber\\
&=& Z_b^{-1}  \sum_{(x,y)} \Delta^{-1}  \min \{w(y),w(x)\}
\nonumber\\
&=& Z_b^{-1} Z_{\{b,c\}}
\text{ and thus}\\
q_{c \rightarrow b} &=& Z_c^{-1} Z_{\{b,c\}} ~.  \label{eq:qcb}
\end{eqnarray}

Equation (\ref{eq:qbc}) considers all microscopic transitions $x\rightarrow
y$ from a micro-state $x$ in $b$ to a micro-state $y$ in $c$, based on the
probability of $x$ $(P_b(x))$ and the transition probability $p_{x
  \rightarrow y}$. The energetically higher micro-state of each such
transition contributes to the partition function of all transition states
between $b$ and $c$, $Z_{\{b,c\}}$ (Eq.~\ref{eq:qbc}
and~\ref{eq:qcb}). Consequently $Z_{\{b,c\}} \equiv Z_{\{c,b\}}$,
\ie{} the transition state partition function is direction-independent.

Within this work, we use the common gradient basin partitioning of $X$
following \change{\citep{Doye:02,Flamm:02,Flamm:08a,Mann:Klemm:11}}. A gradient
basin is defined as the set of all states who have a steepest descent (gradient)
walk ending in the same local minimum, where $\check{x}$ is a local minimum
if $\forall_{y\in \ELneighX(\check{x})} : E(\check{x}) < E(y)$. In this
context the set of macro-states $B$ is given by the set of all local minima
of the landscape, whose number is drastically smaller than that of all
micro-states~\citep{Lorenz:11}. The mapping function $F(x)$ applies a
gradient walk starting in $x$, thus assigning it a local minimum
$\check{x}$ and a macro-state $b$. Here, the minimum is used as a
representative for the macro-state comprised of the gradient basin.

\change{A coarse abstraction of the macro-state transition probabilities
can be obtained by an Arrhenius-like transition
model~\citep{Wolfinger:04a}. Here, the transition probability is
dominated by the minimal energy barrier that needs to be traversed in
order to go from one state to another. Formally, given two states $x$ and
$y$, one has to identify the path $p=(x_1,\ldots,x_l)\in X^l,l>1$ with
$x_1=x$, $x_l=y$, and $\forall i<l: x_{i+1}\in\ELneigh{x_i}$ with lowest
energy maximum. Arrhenius barrier-based transition probabilities are thus
defined by}
\begin{eqnarray}
a_{x \rightarrow y} &=& A\exp(-\beta (E(x,y)-E(x))) \text{ with}
\label{eq:abc}\\ 
E(x,y) &=& \min_{p\in X^{\ast}}\max_{x_i\in p}(E(x_i))
\end{eqnarray}\change{where $A$ is an intrinsically unknown pre-exponential factor. For
macro-state transitions based on a gradient basin partitioning, transition
probabilities can be approximated by Arrhenius probabilities among local
minima of macro-states. In this context it is important to note that this
transition model does not enforce neighborhood of the macro-states. The
impact on modeling quality of such an Arrhenius-based model is evaluated
in Sec.~\ref{sec:RNA}.} We will now present approaches for the exact
determination of the macro-state transition probabilities for a given
landscape and partitioning.

\section{Macro-state transition probabilities}

Following the rationale presented above, all macroscopic transition rates
need to be determined in order to study the coarse-grained dynamics.  Given
Eq.~\ref{eq:qbc}, the partition function~$Z_b$ (Eq.~\ref{eq:zb}) and
adjunct partition functions of transition states $Z_{\{b,c\}}$ to adjacent
$c\neq b$ have to be computed for each macro-state $b$.

A direct approach is brute-force enumeration of $X$, computing $F(x)$ for
each micro-state $x\in X$ and updating $Z_{F(x)}$ accordingly.
Subsequently, all neighbors $y\in\ELneigh{x}$ are enumerated in order to
update $Z_{\{F(x),F(y)\}}$ if $F(x) \neq F(y)$. While this is the simplest and
most general approach, it is not efficient for the majority of
definitions of~$F$. It can, however, be replaced with more
efficient dedicated flooding algorithms and can be even more tuned for
gradient basin definitions of~$F$ as we will discuss now.

\subsection{Standard approach via global flooding}

\newcommand{\setDone}{\textbf{D}}
\newcommand{\setToDo}{\textbf{T}}

The \lid{} method \citep{SchoenSibani:98,Sibani:99} performs a
``spreading'' enumeration starting from a local minimum with an upper
energy bound for micro-states to consider, the lid. Internally, two lists
are hashed: The set \setDone{} containing all micro-states that have been
processed so far and the ``todo-list'' \setToDo{} comprised of states
neighbored to \setDone{} but not handled yet. Each processed
micro-state~$x$ is assigned to its corresponding macro-state $b = F(x)$
during the enumeration process. $b$ is stored along with $x$ in \setDone{}
and \setToDo{} and the partition function $Z_b$ is updated by $w(x)$
accordingly. Subsequently, all neighbors $y\in \ELneigh{x}$ of~$x$ with
$E(y) <$ lid-threshold are enumerated and either found in \setDone{} or
\setToDo{} (thus saving $F(x)$ computation) or added to \setToDo{}. If the
macro-state assignment for $x$ and $y$ differs, \ie{} $F(x)\neq F(y)$, the
corresponding transition state partition function $Z_{\{F(x),F(y)\}}$ is
increased by $\Delta^{-1}\min(w(x),w(y))$. The method was reformulated by
\citet{Kubota05} for DNA energy landscapes and \citet{Wolfinger:06a} in the
context of lattice proteins.

The \barriers{} approach by \citet{Flamm:02} performs a ``bottom-up''
evaluation of energy landscape topology based on an energy-sorted list of
all micro-states in $X$ above the ground state up to a predefined energy
threshold. Here, the macro-state assignment $F$ can be handled more
efficiently compared to the \lid{}-method if gradient basins are applied:
Given that the steepest descent walk used for a gradient mapping $F$ is
recursive, \ie{} the assignment $F(x)$ of a state $x$ is known as soon as
the assignment of its steepest descent neighbor $m_{\min} \in \ELneigh{x}$,
$F(m_{\min})$, is known, the macro-state assignment is accomplished by a
single hash lookup: Since the processed set of states \setDone{} already
contains all states with energy less than $E(x)$, looking up $m_{\min}$ and
its corresponding macro-state $F(m_{\min})$ in \setDone{} yields
$F(x)\equiv F(m_{\min})$. The energy of the micro-state currently processed
marks the ``flood level'', \ie{} all states in $X$ with energy below have
been processed. Consequently, the macro-state partition functions $Z_b$ are
collected as soon as the flood level reaches the according local minimum
defining~$b$.

Both methods perform a massive hashing of processed states and are thus
restricted by memory, \ie{} the number of micro-states that can be stored
in \setDone{} and \setToDo{} is constrained to the available memory
resources. Considering the exponential growth \eg{} of the RNA structure
space $X$ \citep{Hofacker:1998}, the memory is easily exhausted for
relatively short sequence lengths. As the memory limit is approached, both
methods result in incomplete macro-state transition data.

The \barriers{} approach ensures a ``global picture'' of the landscape
since it covers the lower parts of all macro-states up to the reached flood
level exhaustively, missing all macro-states above the limit. In case the
transition states connecting the macro-states are above the flood level, no
transition information is available. This can be approached by heuristics
approximating the transition barrier
\citep{Morgan:1998,Flamm:00b,Wolfinger:04a,Richter:07,Bogomolov:10}, however the outcome
is still not reflecting the true targeted macro-state dynamics. In
contrast, the \lid{} method will always result in connected macro-states
but only a restricted part of the landscape is covered. Furthermore, each
macro-state is enumerated up to different (energy) heights resulting in
varying quality of the collected partition function estimates, which
further distorts the dynamics.

\subsection{Memory efficient local flooding}

To overcome the memory limitation of global flooding approaches, we
introduce a local flooding scheme. It enables parallel identification of
the partition function $Z_b$ and all transition state partitions
$Z_{\{b,c\}}$ for a macro-state $b$ without the need of full landscape
enumeration.

\change{Similar} to global flooding, the \emph{local} approach uses a set
\setDone{} of already processed micro-states that are \emph{part of $b$}, \ie{} $\forall_{x\in
\setDone{}} : F(x) = b$, and a set \setToDo{} of micro-states that might be part
of $b$ or adjacent to it.

The algorithm starts in the local minimum $l\in X$ of $b$, \ie{} $F(l) = b$ and
$\forall_{x\neq l \in F^{-1}(b)} : E(x) > E(l)$, and does a local
enumeration of micro-states in increasing energy order starting from $b$. Thus,
$Z_b$ is initialized with $Z_b = w(l)$, all neighbors $m\in\ELneigh{l}$ of the
minimum are pushed to \setToDo{}, and $l$ is added to \setDone{}. Afterwards the
following procedure is applied until \setToDo{} is empty.

\begin{enumerate}
  \item get energy minimal micro-state $x$ from \setToDo{} with $\forall_{x'
  \neq x \in \setToDo{}} : E(x) < E(x')$
  \item identify steepest descent neighbor $m_{\min} \in \ELneigh{x}$ with $\forall_{m \neq m_{\min} \in
  \ELneigh{x}} : E(m_{\min}) < E(m)$
  \item if $m_{\min} \in \setDone{}$ $\rightarrow F(x) = b$ : \begin{itemize}
	    \item $x$ is added to \setDone{},
	    \item $Z_b = Z_b + w(x)$,
	    \item all neighbors $m \in \ELneigh{x}$ with $E(m) > E(x)$ are added to
	    \setToDo{}, and
	    \item descending transitions leaving $b$ are handled:\\
	    $x$ is transition
	    state for all $m \in
	    \ELneigh{x}$ with $E(m) < E(x)$ and $m \not\in \setDone{}$ :\\
	    $Z_{\{b,F(m)\}} = Z_{\{b,F(m)\}} + \Delta^{-1}w(x)$
	  \end{itemize}
	  else $\rightarrow F(x) \neq b$: \begin{itemize}
	    \item descending transitions entering $b$ are handled:\\
	    $x$ is transition
	    state for all $m \in
	    \ELneigh{x}$ with $E(m) < E(x)$ and $m \in \setDone{}$ :\\
	    $Z_{\{F(x),b\}} = Z_{\{F(x),b\}} + \Delta^{-1}w(x)$
	  \end{itemize}
\end{enumerate}

We use a data structure for \setToDo{} that is automatically sorted by
increasing energy in order to boost performance of step~1.

The algorithm computes $Z_b$ and $Z_{\{b,c\}}$, which are required for
deriving the macro-state transition rates $q_{b\rightarrow c}$
(Eq.~\ref{eq:qbc}) from one macro-state~$b$ to adjacent macro-states $c\neq
b$. It is individually applied to all macro-states in order to get the full
transition rate information of the energy landscape. Evidently, the
transition state partition function $Z_{\{b,F(x)\}}$, covering states
between two macro-states $b$ and $c$, has to be computed only once for each
pair (see Eq.~\ref{eq:qbc} and \ref{eq:qcb}).

The major advantage of the local flooding method compare\change{d} to
global flooding approaches is an extremely reduced memory consumption. This
is achieved by only storing the micro-states part of the current
macro-state $b$ (set \setDone{}) plus all member and transition state
candidates (set \setToDo{}).  The reduction effect is studied in detail in
the next section and an implementation \change{of the presented local
  flooding has been added to the} Energy Landscape Library (ELL)
\citep{Mann_ELL_BIRD07}.  \change{The ELL provides a generic platform for
  an independent implementation of algorithms and energy landscape models
  to be freely combined \citep{Mann_2008_2,Mann:Klemm:11}. Within this
  work, we tested our new method using the ELL-provided RNA secondary
  structure model as discussed in the following section.}

The reduced memory consumption of the local flooding scheme comes at the
cost of increased computational efforts for the assignment of states that
are not part of macro-state $b$. The above workflow does an explicit
computation of $F$ for all these states. Here, more sophisticated methods
can be applied that either do a full or partial hashing of states partaking
in steepest descent walks to increase the performance.

Another advantage is the inherent option for distributed computing since
the local flooding is performed independently for each macro-state. As
such, we can yield a highly parallelized transition rate computation not
possible in the global flooding scheme. This can be combined with an
automatic landscape exploration approach where each local flooding instance
identifies neighboring, yet unexplored macro-states that will be
automatically distributed for processing until the entire energy landscape
is discovered.

We will now investigate the requirement and impact of our local flooding
approach in the context of folding landscapes of RNA molecules.

\section{Folding landscapes of RNA molecules}
\label{sec:RNA}

In the following, we will study the energy landscapes for the bistable RNA
d33 from \citep{Mann:Klemm:11} and the iron response element (IRE) of the
Homo sapiens L-ferritin gene (GenBank ID: KC153429.1) in detail. The
sequences are
\texttt{GGG\-AAU\-UAU\-UGU\-UCC\-CUG\-AGA\-GCG\-GUA\-GUU\-CUC} and
\texttt{CUG\-UCU\-CUU\-GCU\-UCA\-ACA\-GUG\-UUU\-GGA\-CGG\-AACAG},
respectively. In addition, and in order to evaluate the general character
of our results, we generated 110 random RNA sequences with uniform base
composition, 10 for each length from 25nt to 35nt. For this set average
values are reported. The length restriction was a requirement for
comparison to exhaustive methods.

\subsection{Exact vs. approximated transition models}

We will first investigate whether exact macro-state transition
probabilities are essentially required for computing a coarse-grained
dynamics or if an approximated model is providing similar results. To address
this question, we performed an exhaustive enumeration of the RNA energy
landscapes for d33 and ire, resulting in approximately 30 and 21 million
micro-states, respectively, that are clustered into approximately 2,900
gradient basin macro-states for each sequence. These basins are connected
by approximately 60,000 macro-state transitions, representing only a
fraction of 1.5\% of all possible pairwise transitions.

The concept of barrier trees \citep{Flamm:02,Flamm:08a} represents a
straightforward approach for modelling the coarse-grained folding dynamics of an
RNA molecule without explicit knowledge of the exact pair\-wise microscopic
transition probabilities. In this context, transition probabilities between any
two gradient basin macro-states $b$ and $c$ are defined via an Arrhenius-like
equation. The latter is given \change{in Eq.~\ref{eq:abc}}, considering the
energy difference $\Delta E$ between the local minimum of macro-state $b$ and
the lowest saddle point of any path to the target macro state $c$ (which may
traverse some other macro-states). The saddle point can be identified either via
exhaustive enumeration \citep{Flamm:02} or estimated by path sampling techniques
\change{\citep{Richter:07,Lorenz:09,Bogomolov:10,Li:12,Kucharik:2014}}. Energy
barriers can be visualized in a tree-like hierarchical data structure, the barrier tree,
resulting in all $n^2$ pairwise transition probabilities for $n$ macro-states.
Coarse-grained folding kinetics based on this framework have been shown to
resemble visual characteristics of the exact macro-state kinetics
\citep{Flamm:02,Wolfinger:04a}.

\change{The supplementary material} provides a visual
comparison of coarse-grained folding dynamics for RNA d33, based on two
different transition models. While the pure barrier tree dynamics resembles the
overall dynamics of the two energetically lowest macro-states of the exact model
quite well, it shows significant differences for states populated at lower
extent. Given these visual discrepancies, we are interested in measuring the
modelling quality of the barrier tree-based transition model vs. the exact
configuration. To this end, we will analyze mean first passage times (FPT) and
their correlations. The FPT $\FPT(b,t)$, also termed exit time
\citep{Freier:86}, is the expected number of steps to reach the target state
$t\in B$ from a start state $b\in B$ for the first time \citep{Grinstead:97}.
The first passage time for a state to itself is $0$ per definition, \ie{}
$\FPT(b,b)=0$. For all other cases, it is defined by the recursion

\begin{equation}
\FPT(b,t) = 1 + \sum_{c\in B} q_{b\rightarrow c}\FPT(c,t).
\end{equation}
We are focused on folding kinetics, \ie{} we compute the FPT from the
unfolded state to all other macro-states using (a) the exact macro-state
transition probabilities (Eq.~\ref{eq:qbc}) (full model) and (b) the
barrier tree-based transition probabilities based on the Arrhenius
equation \change{(Eq.~\ref{eq:abc}, barrier model)}.

First passage time values depend on the intrinsically unknown Arrhenius
prefactor. As such, we will compare the two models using a Spearman rank
correlation of the FPT, \ie{} we compare the relation between FPTs rather
than final values.

\begin{table}   
\begin{center}
\footnotesize
\begin{tabular}{|c|c|c|}
\hline
  Sequence(s)  & Spearman corr. & Spearman corr.\\
	 & \change{exact} -- barrier   	
	 & \change{exact} -- merged \\
\hline
d33    & 0.28	  & 0.85  \\   
ire	   & -0.12  & 0.64 \\
random & 0.20   & 0.71 \\
\hline
\end{tabular}
\end{center}
\caption{\small Spearman rank correlation of different \change{macro-state
transition} models. Comparison of the \change{Arrhenius barrier-based and
the exact model (Eq.~\ref{eq:qbc})} shows almost no
correlation, while the merged model \change{of both (see text)} is highly
correlated to the \change{exact} model.}
\label{T:spearman}
\end{table}

For d33 and ire the Spearman rank correlation coefficients is 0.28 and
-0.12, respectively, indicating no correlation. The random sequence set
shows a mean coefficient of 0.2, indicating no correlation either. \change{No
length-dependent bias was found (see suppl. material).} Results are
summarized in Table \ref{T:spearman}.

The barrier model is a simplification of the full model in two aspects: 1.)
\emph{loss of precision} -- the computation of transition rates based on
Arrhenius-like equations is less accurate and 2.) \emph{loss of topology}
-- the barrier model allows for all possible pairwise transitions, which
may lead to an overestimation of transitions. To further distinguish
between these two transition approaches, we have derived a merged
transition model with modified transition probabilities $q'$. Within this
\change{merged} model, $q'_{b\rightarrow c}$ is given by the Arrhenius-like
equation \change{(Eq.~\ref{eq:abc})} for all exact macro-state transitions
($q_{b\rightarrow c} \neq 0$\change{, Eq.~\ref{eq:qbc}}) and zero otherwise.
Investigating the Spearman rank correlation of the merged model's FPTs with the
exact FPTs, an increased correlation coefficient (0.85 and 0.64 for d33 and ire,
resp.), is observed. This is supported by a \change{robust} average coefficient
of 0.71 for the set of random sequences \change{(see suppl. material)}.

These results clearly show two key aspects of reduced folding
dynamics: First, importance of the underlying topology of the landscape,
\ie{} the necessity to identify sparse exact transitions between
macro-states, and second the reduced modeling quality when restricting the
computation of transition probabilities to energy barrier-based
(Arrhenius-like) approximations. The importance of the topology information
for kinetics is partly studied in the supplementary material of
\citet{Kucharik:2014}.

\subsection{Reduction of memory requirement}

Given the need for an exact computation of macro-state transition
probabilities, we will now evaluate the impact of a local flooding scheme
compared to the standard global flooding approach. In this context, we will
investigate the memory footprint, which is the central bottleneck of global
flooding methods.

\newcommand{\mem}[1]{\text{mem(#1)}}

As outlined above, global flooding schemes keep track of all micro-states
$x\in X$ within the energy landscape. As such, the global flooding
memory consumption is dominated by $\mem{G} = |X|$.

In contrast to that, all micro-states $x\in F^{-1}(b)$ of $b$ in the local
flooding scheme have to be stored in order to compute $Z_b$
(Eq.~\ref{eq:zb}) as well as the set of all micro-state transitions leaving
macro-state $b$, denoted $T(b)$, for computing $Z_{\{b,\ast\}}$
(Eq.~\ref{eq:qbc}). The memory consumption of local flooding of $b$ is thus
ruled by $\mem{L}=|F^{-1}(b)|+|T(b)|$.

\begin{figure}
\begin{center}
  \includegraphics[width=0.8\columnwidth]{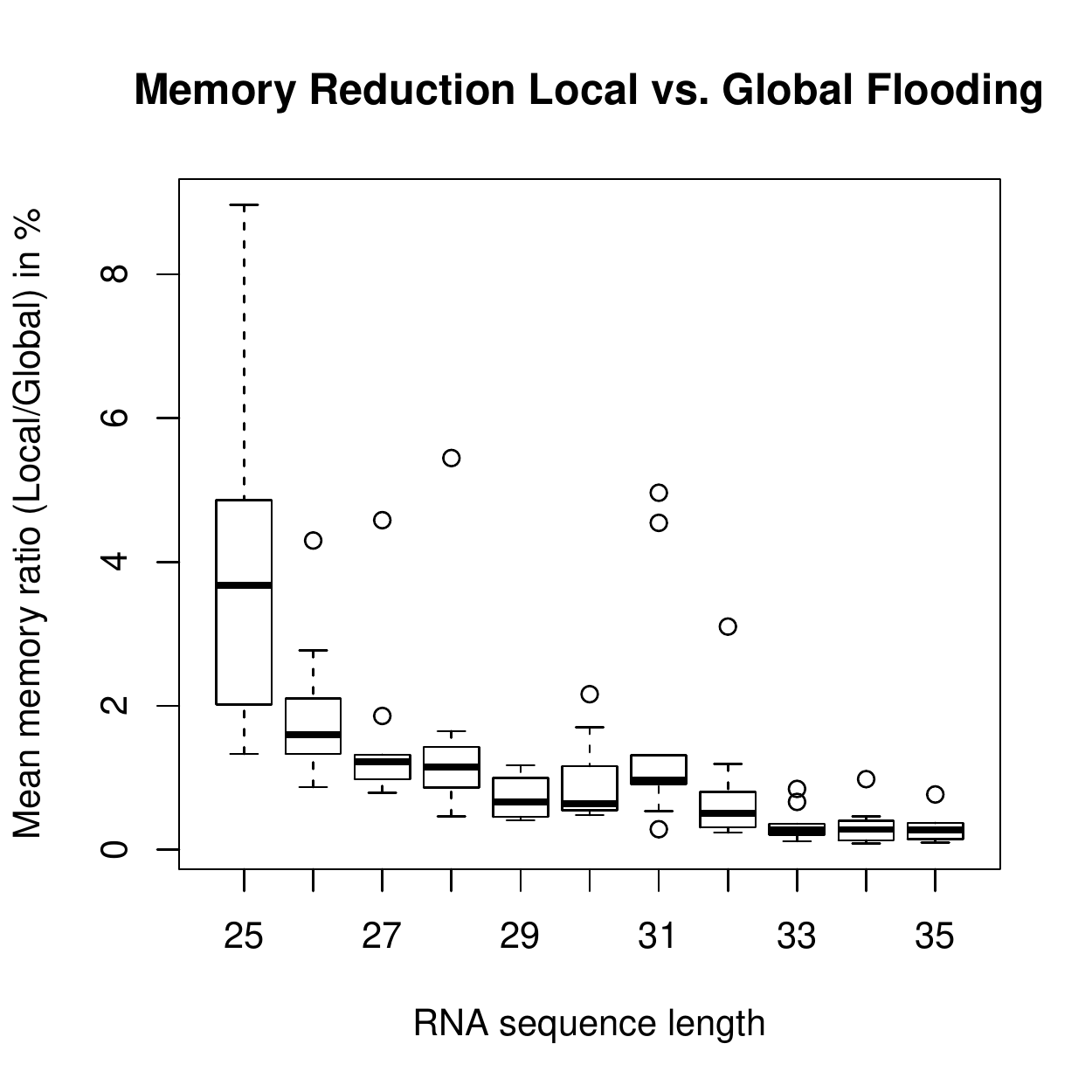}
  \vspace{-1em}
  \caption{Memory consumption comparison of local vs. global flooding for the
  random sequence set. For each RNA sequence length, 10 mean ratios of local vs.
  global flooding memory requirement are measured and visualized in a box plot.
  The box covers 50\% of the values and shows the median as horizontal bar. A
  similar picture is obtained when plotting the mean gradient basin size for
  each sequence.}
  \label{fig:memRandomSet}
\end{center}
\end{figure}

Investigating the ratio of $\mem{L}/\mem{G}$ for all macro-states, we find
a mean value of $0.0015$ and a median of $< 0.0001$ for both the d33 and
the ire landscape. In other words, the memory footprint of local flooding
comprises less than $0.005$ (0.5\%)
compared to global flooding for almost all macro-states (99\%). For
approximately 80\% of the macro-states, the footprint drops even lower to
less than 0.01\%. Similar numbers are observed within the random set for
sequences of same lengths. Most notably, we see a logarithmic decrease of
the average memory reduction with growing sequence length (see
Fig.~\ref{fig:memRandomSet}). We find only three large macro-states with
$\mem{L}/\mem{G} > 10$\% in both landscapes.

These numbers give evidence for the memory efficiency of a local flooding
scheme. Within the context of extensive parallelization, such a scheme can
be applied to large energy landscapes, since the individual memory
consumption is several orders of magnitudes lower compared to a global
flooding scheme. The remaining set of few large macro-states can be handled
at the cost of longer runtimes by using the efficient local sampling scheme
for macro-state transition probabilities presented
in~\citep{Mann:Klemm:11}.

\subsection{Properties of gradient basins}

In the following, we will work out various properties of gradient basins,
since they are commonly used as macro-state abstraction in RNA energy
landscapes. We will give examples for RNA d33, however the results can be
generalized to other RNAs as shown for the random sequence set.

\begin{figure}
\begin{center}
\includegraphics[width=0.8\columnwidth]{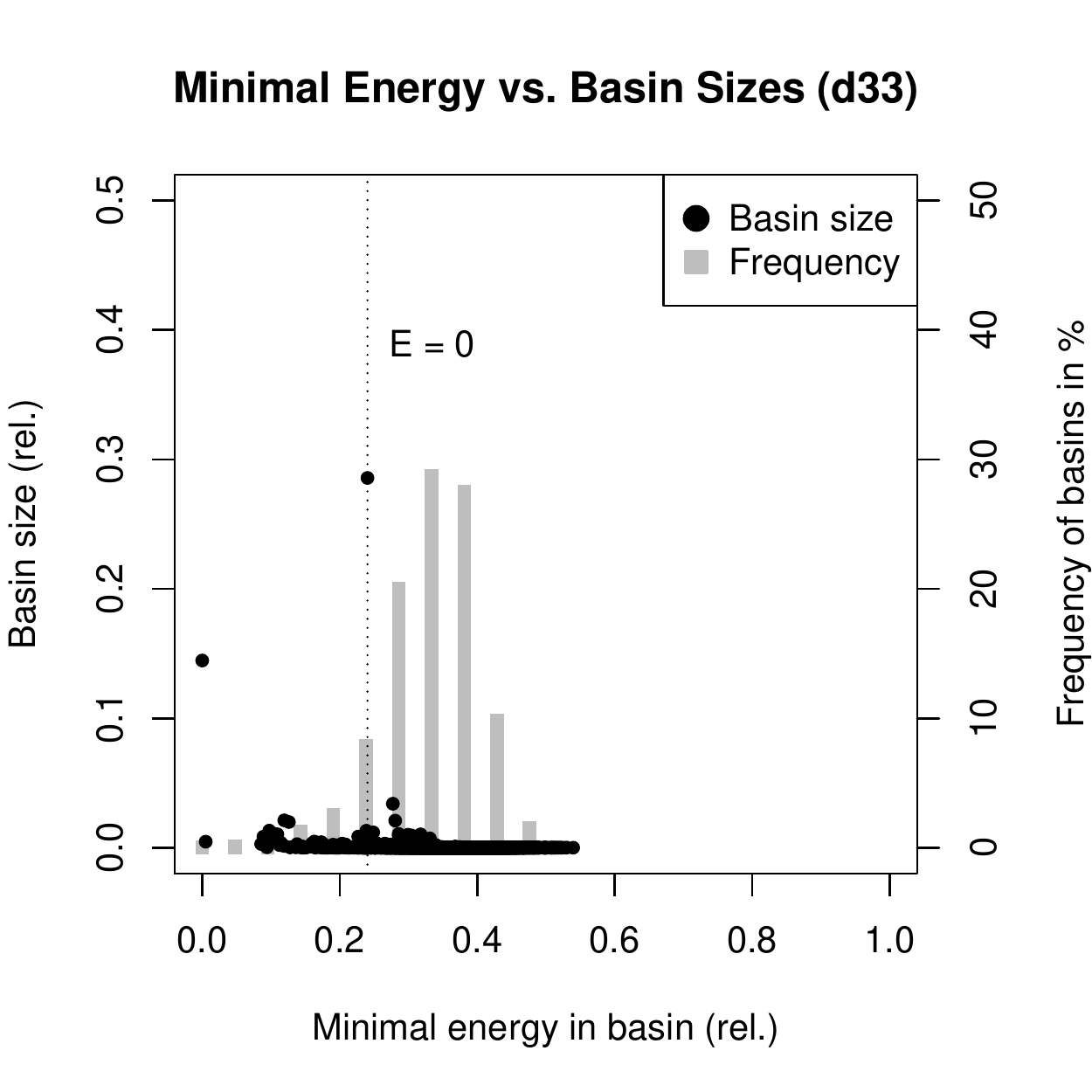}
\vspace{-1em}
  \caption{\label{fig:basinSize} Distribution of basin sizes (dots) and
  frequency histogram of basins (bars) over the energy range within the
  energy landscape of RNA d33. \change{Relative energies are given by
  $E_{\text{rel}}(x) = (E(x)-E_{\min})/(E_{\max}-E_{\min})$ where
  $E_{\min}$/$E_{\max}$ denote the energy boundaries over $X$.} The dotted
  line marks the position of the unstructured state with energy~0.}
\end{center}
\end{figure}

We \change{have shown} in the context of local flooding memory consumption that
the overwhelming majority of gradient basins is small, whereas there are only a
few densely populated gradient basins. Most importantly, the basin of the
open, unstructured state, which is a local minimum according to the Turner
energy model \change{\citep{xia:98}} and the selected neighborhood relation
$\ELneighX$ allows for the largest neighborhoods. Consequently, its gradient basin is the largest
for all RNAs studied and wraps about 20-30\% of the state space. In the
random data set, the open state covers on average approximately 40\% of the
landscape and we see a decrease of this fraction with increasing sequence
length. The same tendency applies to the average relative basin size (see
Fig.~\ref{fig:memRandomSet}). Other large gradient basins are usually
centered at energetically low local minima and their basin size is in
general highly specific for the underlying sequence. \change{We do observe a
correlation of basin size with the energy of its local minimum (Spearman 
corr. -0.73), which is in accordance to the findings of
\cite{Doye:98} for Lennard-Jones clusters.}

When investigating the distribution of the energetically lowest
micro-states in each gradient basin, \ie{} the local minima, we find that
most minima have positive energies (see histogram in
Fig.~\ref{fig:basinSize}). Minima are distributed over the lower 40-50\% of
the energy range for all sequences studied. The number of minima with
negative energy, \ie{} stable secondary structures, is approximately 100 for
d33 and ire and is in the range of approximately 5\% in general for the
random set studied here. The majority of the state space of RNA energy
landscapes shows positive energies, resulting from de-stabilizing energy
terms for unstacked base pairs in the Turner energy model
\change{\citep{xia:98}}. This is in accordance with the results from \citet{Cupal_1997} who found that only
$\sim 10^{6}$ of $\sim 10^{16}$ structures of a tRNA show an energy of less
than zero.

\begin{figure}
\begin{center}
\includegraphics[width=0.8\columnwidth]{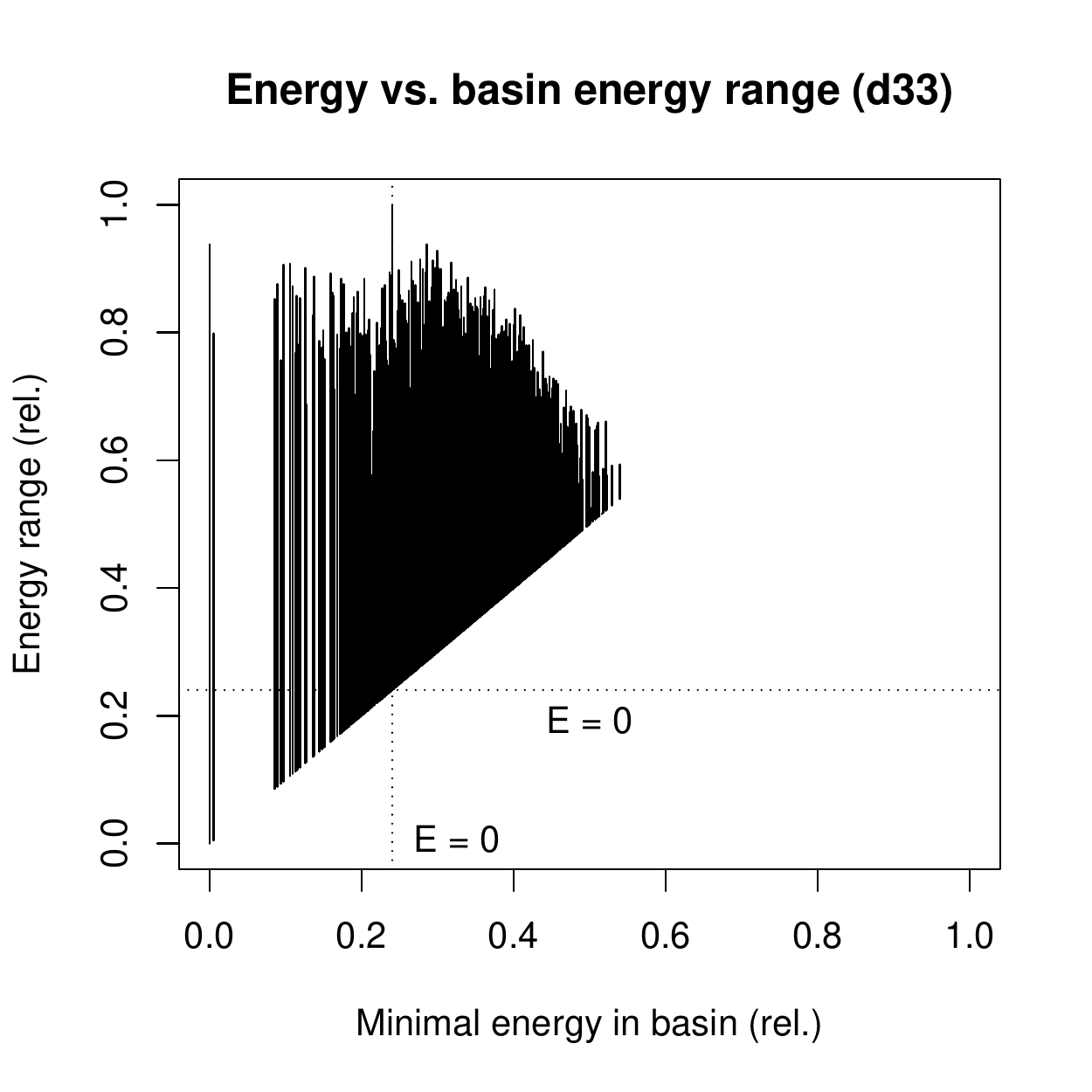}
\vspace{-1em}
  \caption{\label{fig:energyRange} The energy range covered by each
basin (Y-axis) sorted by the minimal energy within the basin (X-axis) over the
whole energy range of the energy landscape of RNA d33. \change{Relative energies
are given by $E_{\text{rel}}(x) = (E(x)-E_{\min})/(E_{\max}-E_{\min})$ where
$E_{\min}$/$E_{\max}$ denote the energy boundaries over $X$.} The dotted lines
mark the position of the unstructured state with energy~0. }
\end{center}
\end{figure}

The energy range of most gradient basins, \ie{} minimal to maximal energy of any
micro-state in the basin as plotted in Fig.~\ref{fig:energyRange}, covers almost
the entire range above a local minimum. This is generally independent of the
basin size (compare Fig.~\ref{fig:basinSize} and \ref{fig:energyRange}), only
for energetically high basins a lower maximal energy is observed. \change{This
might be a result of the accompanying basin size decrease or an artifact of the
energy model}. The gradient basin of the unstructured state covers the
energetically highest states.

As mentioned above, only few of the possible $|B|^2$ macro-state transitions are
observed. We find that more than 50\% of the basins show less than 10
neighboring basins and almost all (98\%) have transitions to less than 2\% of
the basins. The gradient basin of the unstructured state shows the highest
number of macro-state transitions and is connected to more than 20\% of the
macro-states. We find that few large basins serve as hub nodes with high
connectivity. \change{This is in accordance to findings of \cite{Doye:02} for
Lennard-Jones polymers.} Consequently, the number of transitions is highly
correlated to the basin size, as one would expect. This is supported by a
Spearman rank correlation coefficient of approx. 0.8 for all RNAs studied.
\change{The correlation to the basin's minimal energy, as found by
\cite{Doye:02}, is not as significant (Spearman corr. -0.6).}

\vspace{1em}
\section{Conclusion} \label{sec:concl}

We have introduced a local flooding scheme for computing the exact
macro-state transition rates for arbitrary discrete energy landscapes
provided some macro-state assignment is available. The approach has been
evaluated on RNA secondary structure energy landscapes in the context of
modeling coarse-grained RNA folding kinetics based on gradient basins. We
have demonstrated the need for exact macro-state transition models via
comparison to a simpler, barrier tree-based Arrhenius-like
model. The latter resulted in significantly different dynamics measured by
mean first passage times.

We \change{showed} that the local flooding scheme requires several orders of
magnitude less memory compared to the standard global flooding scheme.
Furthermore, it is intrinsically open to vast parallelization, which should
also result in significant runtime reduction, given that the global
flooding can not be easily parallelized.

Finally, we performed a thorough investigation of gradient basins in RNA
energy landscapes, since they are commonly used as macro-state abstraction
in the field. Gradient basins have been shown to be generally small, which
is the reason for the tremendously reduced memory requirement of the local
flooding scheme. The basin of the unstructured state has been shown to be
special since it is the largest, most connected macro-state and covers the
energetically highest micro-states. Independent of their size, most basins
contain micro-states of almost the entire energy range above their
respective local minimum. The majority of the gradient basins covers only
states with positive energy. We \change{found} a low average connectivity
between gradient basins, the existence of few highly connected hub nodes,
and a high correlation of connectivity with basin size.

\section*{Acknowledgement}

This work was partly funded by the Austrian Science Fund (FWF) project
``RNA regulation of the transcriptome'' (F43), the EU-FET grant RiboNets
323987, the COST Action CM1304 ``Emergence and Evolution of Complex
Chemical Systems'' and by the IK Computational Science funded by the
University of Vienna.


\bibliographystyle{natbib}
%
%

\bibliography{el-explore}

\appendix

\clearpage
\begin{center}
\LARGE Supplementary Material
\end{center}

\section{Exact vs. approximated transition models}

\begin{figure}[h!]
\begin{center}
  \includegraphics[width=.45\textwidth]{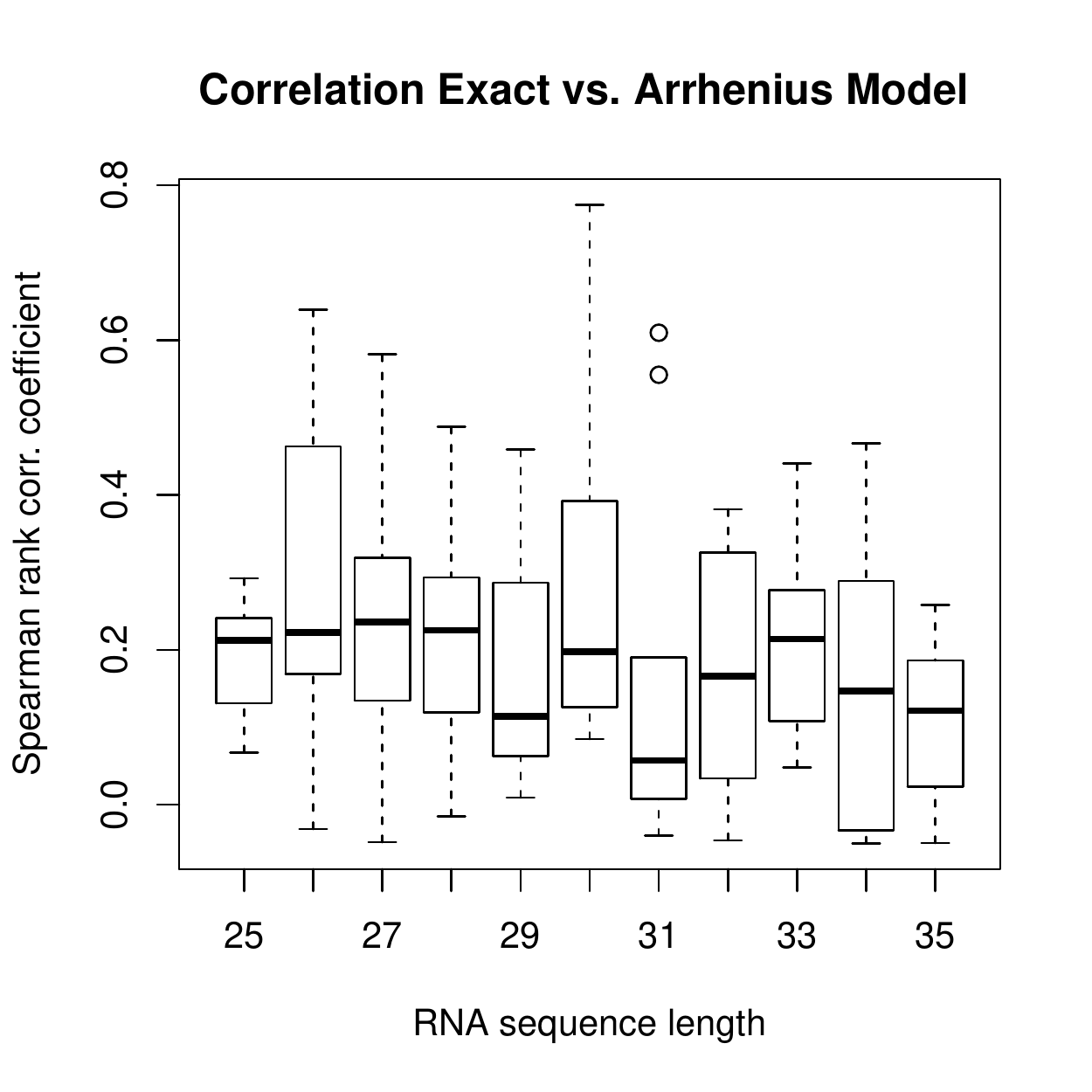}
  \hspace{1em}
  \includegraphics[width=.45\textwidth]{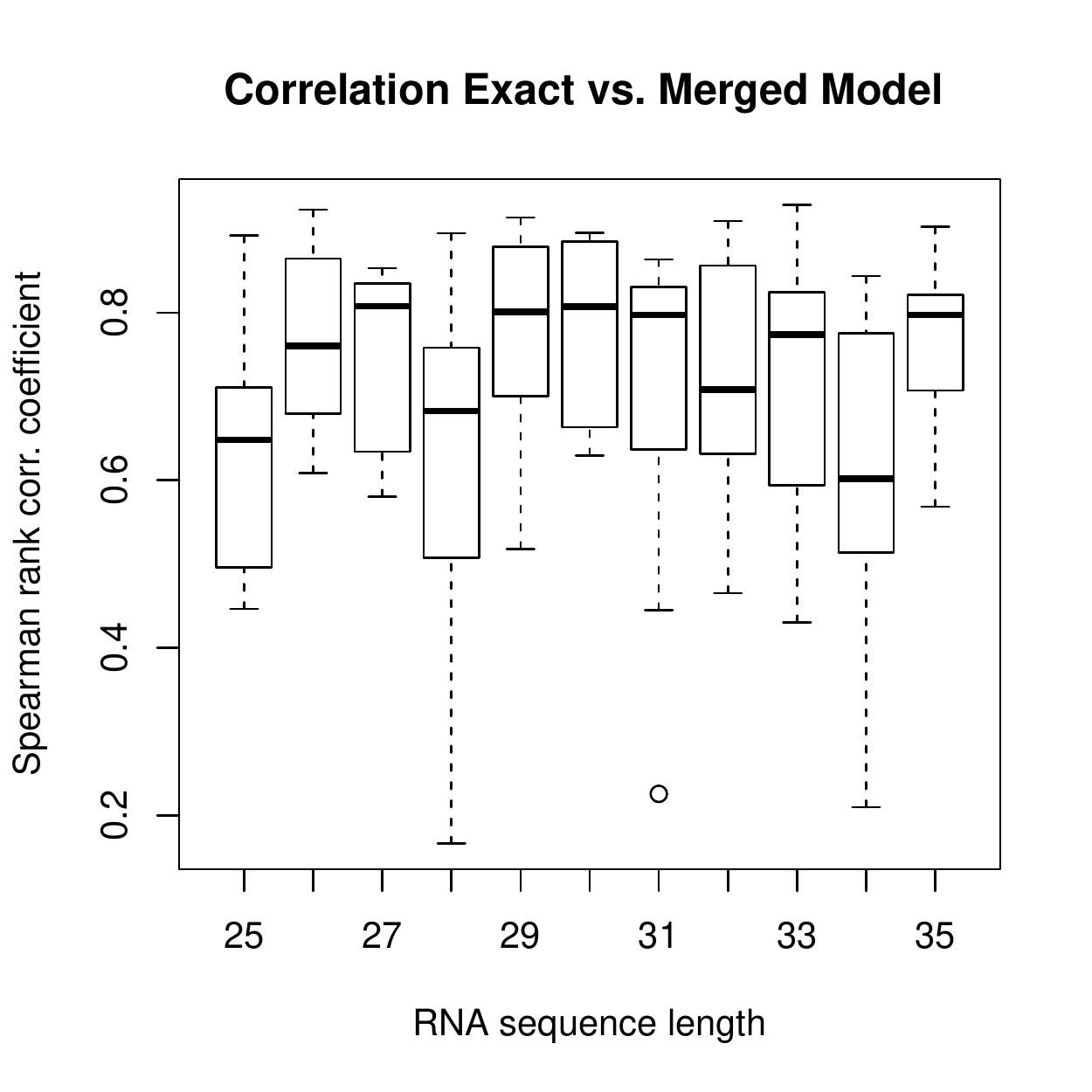}
  \caption{Spearman rank correlation coefficients of the mean first passage
    times (FPT) for the random data set grouped by sequence
    length. Correlation of the exact model (left) with the Arrhenius
    barrier-based transition model (right) and the merged transition
    probability model.}
  \label{fig:corr}
\end{center}
\end{figure}

Figure~\ref{fig:corr} presents the Spearman rank correlation of the mean
first passage times (FPT) for the different transition probability models
studied. The plot is based on the random data set and grouped by sequence
length.

\begin{figure}[h!]
\begin{center}
  \includegraphics[width=.5\textwidth]{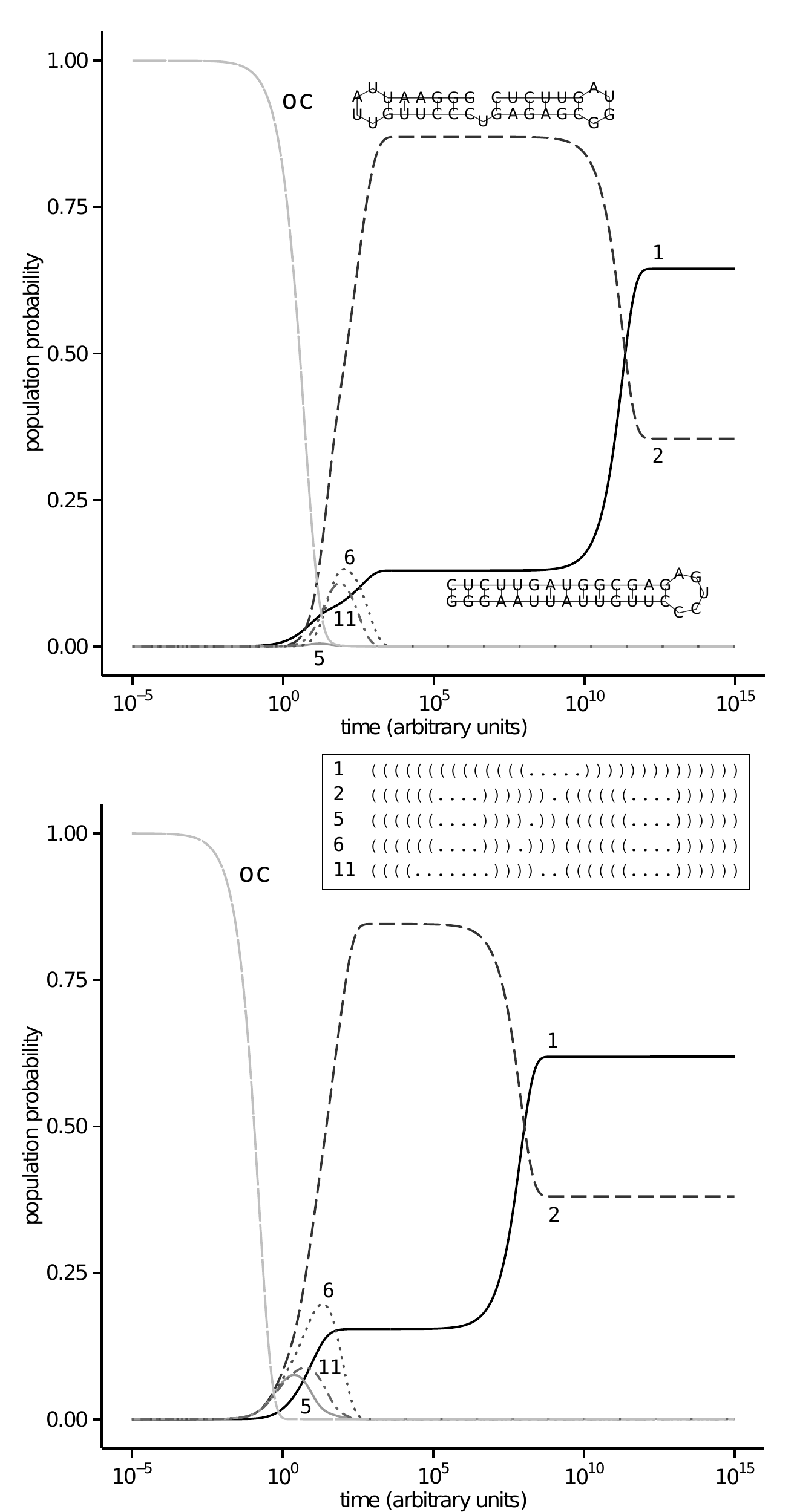}
  \caption{Coarse-grained folding dynamics of RNA d33 showing the five most
    populated gradient basins. Each curve represents the population
    probability of a gradient basin macro state, depicted by the secondary
    structure of its local minimum. Numbers correspond to energy sorted
    ranks. Simulations were started from the unstructured open chain
    macro-state (\texttt{oc} curve) and let evolve until a stationary
    distribution of the underlying Markov process was reached,
    see~\citet{Wolfinger:04a} for details. We compare the dynamics from
    exact transition probabilities (left) to those from a barrier tree-based
    Arrhenius transition model (right).}
  \label{fig:treekin}
\end{center}
\end{figure}

\clearpage
Figure~\ref{fig:treekin} provides a visual comparison of coarse-grained
folding dynamics for RNA d33, based on two different transition
models. While the pure barrier tree dynamics (lower plot) resembles the
overall dynamics of the two energetically lowest macro-states of the exact
model (upper plot) quite well, it shows significant differences for states
populated at lower extent (\eg{} at rank 5 or~6).

\clearpage
\section{Memory Consumption Local vs. Global Flooding}

\begin{figure}[h!]
\begin{center}
  \includegraphics[width=0.45\textwidth]{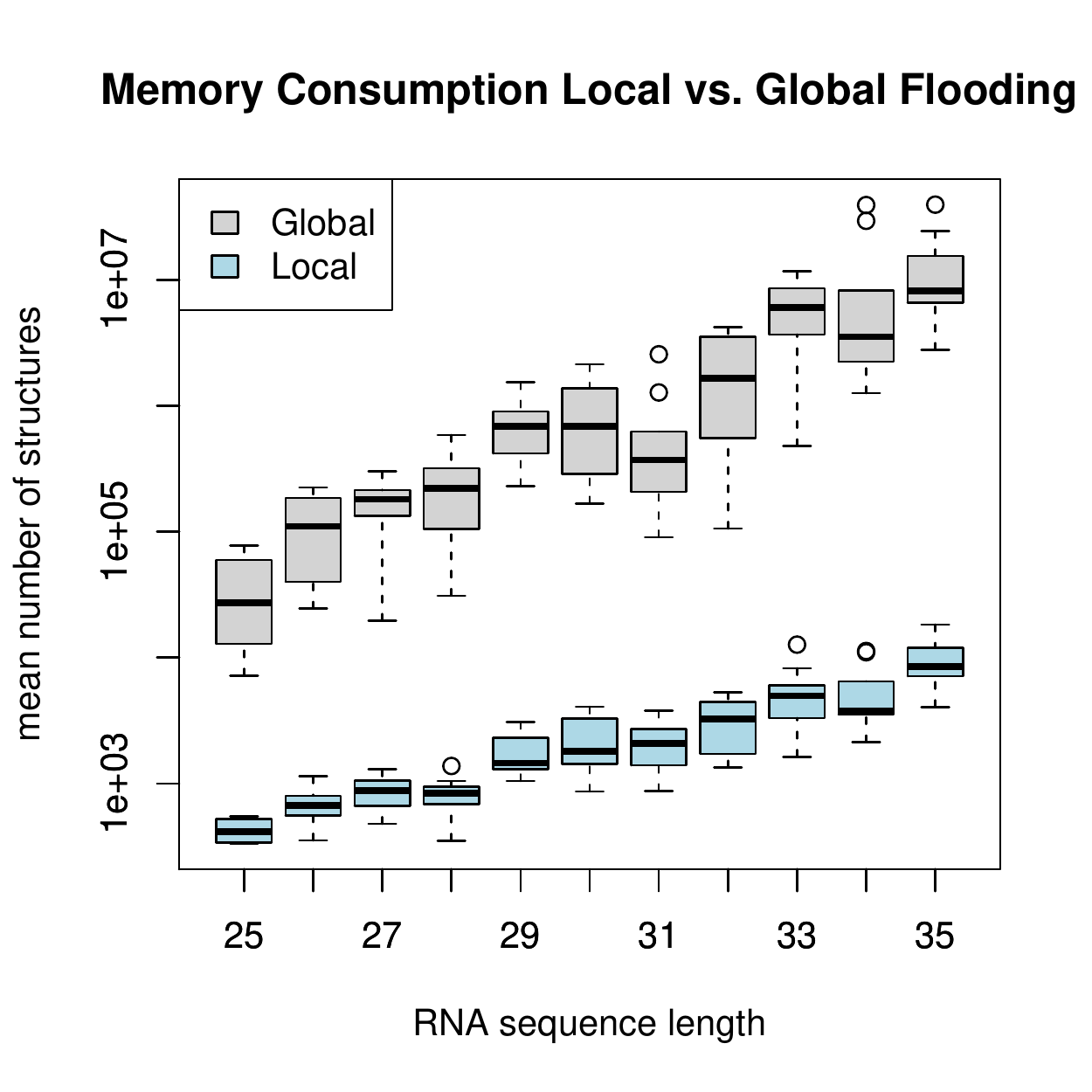}
  \hspace{1em}
  \includegraphics[width=0.45\textwidth]{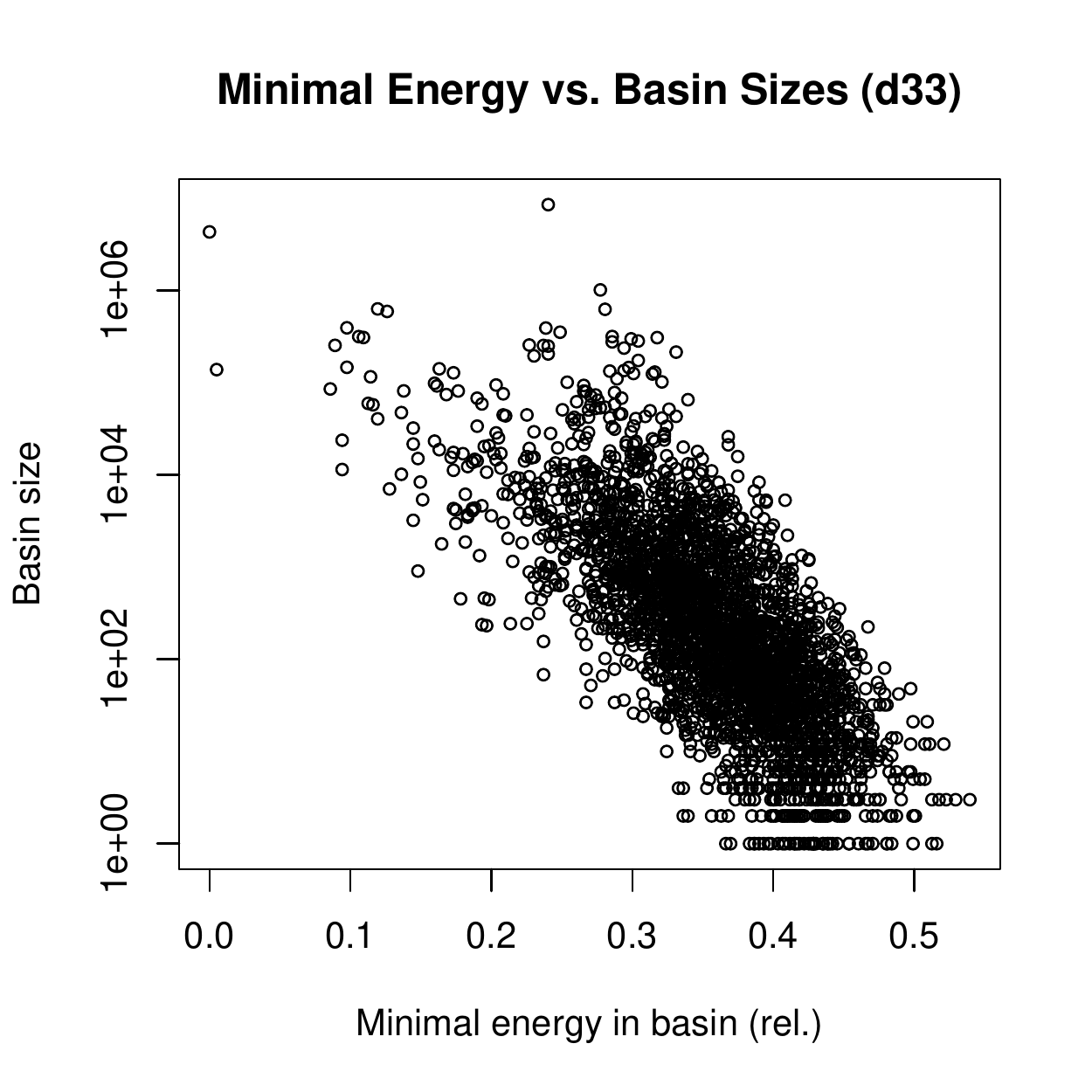}
  \caption{Memory consumption of global and local flooding for different
    RNA lengths within the random data set (left). Distribution of gradient
    basin sizes on a logarithmic scale over the energy range for RNA d33
    (right).}
  \label{fig:mem-consumption}
\end{center}
\end{figure}

In Figure~\ref{fig:mem-consumption} on the left, we present the memory
consumption of the local vs. the global flooding approach in terms of
number of structures to be kept in memory for the random RNA sequence
set. The local flooding requires several orders of magnitude less memory
compared to global flooding. As expected, a growth in sequence length is
visible.

The right side of Figure~\ref{fig:mem-consumption} presents the
distribution of gradient basin sizes over the energy range for RNA d33. A
decrease in basin size is observed with increasing minimal energy. A similar
result was found in the context Lennard-Jones clusters by \cite{Doye:98}.

\end{document}